# Using the kinematic Sunyaev-Zeldovich effect to determine the peculiar velocities of clusters of galaxies

Martin G. Haehnelt[1] & Max Tegmark[1,2]

[1] *Max-Planck-Institut für Astrophysik, Karl-Schwarzschild-Straße 1, D-85740 Garching b. München, Germany*

[2] *Max-Planck-Institut für Physik, Föhringer Ring 6, D-80805 München, Germany*

*e-mail: haehnelt@mpa-garching.mpg.de, max@mppmu.mpg.de*

**Abstract**

We have investigated the possibility of inferring peculiar velocities for clusters of galaxies from the Doppler shift of scattered cosmic microwave background (CMB) photons. We find that if the core radius of the gas distribution or the beam size of the instrument is larger than 3-7 arcminutes, then the maximum attainable signal-to-noise ratio is determined by confusion with primary fluctuations. For smaller angular scales, "cosmic confusion" is less important and instrumental noise and/or foreground emission will be the limiting factor. For a cluster with the optical depth of the Coma cluster and for an optimal filtering technique, typical one-sigma errors span the wide range from 400 to 1600 $\mathrm{km\,s^{-1}}$, depending on the cosmological model, the resolution of the instrument and the core radius of the cluster. The results have important implications for the design of future high-resolution surveys of the CMB. Individual peculiar velocities will be measurable only for a few fast moving clusters at intermediate redshift unless cosmic fluctuations are smaller than most standard cosmological scenarios predict. However, a reliable measurement of bulk velocities of ensembles of X-ray bright clusters will be possible on very large scales ($100 - 500\,h^{-1}$ Mpc).

# 1 Introduction

In a seminal paper, Sunyaev & Zeldovich (1970) discussed the important effect of Compton scattering of cosmic microwave background (CMB) photons by hot electrons. The ionized intra-cluster gas of galaxy clusters leads to a Compton distortion of the CMB behind the cluster on scales of a few arcminutes, and this effect is now generally referred to as the SZ effect (Sunyaev & Zeldovich 1972, 1980a). After a twenty year long history of only tentative detections, the number of convincing observations is now rapidly increasing (Birkinshaw, Hughes & Arnaud 1991; Klein et al. 1991; Grainge et al. 1993; Jones et al. 1993; Birkinshaw & Hughes 1994; Wilbanks et al. 1994). Detector technology and our understanding of the numerous noise sources is developing fast, and it should soon become possible to observe some of the "secondary" effects of this scattering process which were predicted by Sunyaev & Zeldovich (1972, 1980b). The most prominent of these is an additional Doppler shift of the CMB signal due to the radial peculiar motion of the cluster with respect to the CMB rest frame (see Rephaeli & Lahav (1991) for a recent comprehensive discussion). The amplitude of the corresponding CMB distortion is expected to be a factor 10-20 smaller than that for the "ordinary" SZ effect. Following Rephaeli & Lahav (1991), we will refer to the two effects as the *thermal* and the *kinematic* SZ effect, respectively. No convincing detection of the kinematic effect has yet been reported. The best hope for a detection comes from proposed second generation satellite missions, which would reduce pixel noise levels by more than a factor of ten compared to COBE and provide much higher angular resolutions, perhaps ten arcminutes or better — see *e.g.* Mandolesi *et al.* (1995). In addition, a number of ground-based experiments with suitable angular resolution and sufficient sensitivity are currently being built or planned for the near future.

The potential pay-off from a reliable measurement of cluster peculiar velocities with respect to the CMB frame would be enormous. So far, not much is known about bulk velocities on very large scales (redshifts $> 10000\,\mathrm{km\,s^{-1}}$). The measurement of the bulk velocity of a volume-limited sample of 119 Abell clusters out to a distance of $15000\,\mathrm{km\,s^{-1}}$ gave a rather high value of $700 - 800\,\mathrm{km\,s^{-1}}$ (Lauer & Postman 1994; Postman & Lauer 1995; Colless 1995, see Strauss & Willick 1995 for a recent general review on peculiar velocities). This is considerably larger than the approximately $400-500\,\mathrm{km\,s^{-1}}$ expected in most cosmological scenarios currently discussed (Feldman & Watkins 1994; Strauss et al. 1995). An independent measurement method on even larger scales should clarify the situation.

As we will discuss below, on top of technical difficulties and the complications caused by galactic and atmospheric foreground emission, there is an additional important impediment to a measurement of the kinematic SZ effect which so far has received little or no attention: the possible confusion of a Doppler shift of the CMB due to the peculiar velocity of the cluster with primary fluctuations of the CMB on the same scales. This will probably turn out to be a fundamental limitation for a determination of cluster peculiar velocities by the kinematic SZ effect with beam sizes realizable with space missions. In this paper we will review the basic principles of the effect, discuss briefly the numerous problems involved with detecting it, give a detailed assessment of the possible confusion with primary CMB fluctuations and discuss what results future measurements may be able to produce.

# 2 Measuring the kinematic SZ effect

## 2.1 Its strength and spectral signature

Peculiar velocities of the hot intra-cluster gas lead to a Doppler shift of the scattered photons which is proportional to the product of the radial peculiar velocity $v_{\rm pec}$ and the electron density $n_{\rm e}$ integrated along the line of sight through the cluster. For small optical depths, the distortion from a blackbody spectrum is small and the relative change in intensity of the CMB is given by

$$\frac{\Delta I}{I} = \frac{\sigma_{\rm T}}{c} \int_{\rm los} v_{\rm pec}\, n_{\rm e}\, {\rm d}l \left[\frac{x\, e^x}{e^x - 1}\right], \qquad x = \frac{h\nu}{kT}, \tag{1}$$

where $T$ is the temperature of the microwave background and $\sigma_{\rm T}$ is the Thompson cross section (Sunyaev & Zeldovich 1972a,b). In the Rayleigh-Jeans limit $x \ll 1$, this is approximately

$$\frac{\Delta I}{I} \sim \frac{\Delta T}{T} \sim \frac{\sigma_{\rm T}}{c} \int_{\rm los} v_{\rm pec}\, n_{\rm e}\, {\rm d}l. \tag{2}$$

Equation (1) is very similar to the one for the thermal effect, where $v_{\rm pec}/c$ is replaced by $kT_{\rm e}/m_{\rm e}c^2$ and the spectral dependence of the relative intensity change has an extra factor $[x \coth(x/2) - 4]$.

For a typical cluster, the kinematic SZ effect at the cluster center is of order

$$\Delta T \sim 30 \left(\frac{n_{\rm e}}{3 \times 10^{-3}\,{\rm cm}^{-3}}\right) \left(\frac{r_{\rm c}}{0.4\,{\rm Mpc}}\right) \left(\frac{v_{\rm pec}}{500\,{\rm km\,s}^{-1}}\right) \mu{\rm K}, \tag{3}$$

where $n_e$ is the electron density in the core, $r_c$ is the core radius and we have scaled to the values reported by Briel et al. 1991 for the Coma cluster (assuming a distance of 140 Mpc).

## 2.2 Noise sources

The actual observed signal will, however, be a convolution with the beam pattern $B(\mathbf{x})$ of the experiment, and will contain a considerable noise component,

$$\left(\frac{\Delta I}{I}(\mathbf{x})\right)_{\rm obs} = \int B(\mathbf{x}-\mathbf{x}')(S(\mathbf{x}')+N_{\rm sky}(\mathbf{x}'))\,{\rm d}^2\mathbf{x}' + N_{\rm det}(\mathbf{x}), \qquad (4)$$

where the noise-free unconvolved signal $S(\mathbf{x})$ will be given by equation (1). $N_{\rm sky}$ and $N_{\rm det}$ denote the noise in the sky and in the detector, respectively. The area of integration varies considerably between different experiments and can be rather complicated. Most ground-based observations use drift scans and beam switching, while the proposed space-borne missions will scan the whole sky in a complicated pattern. There is a wide variety of noise sources which make a measurement of the kinematic SZ effect rather difficult. The most important are the following:

- Instrumental noise
- Diffuse foreground emission (atmospheric, galactic, extragalactic)
- Radio sources
- The thermal SZ effect of the cluster
- Bulk motion of the gas within the cluster
- Primary CMB fluctuations

In the remainder of this section, we will give a very brief review of the first five of the above noise sources (see Fisher & Lange 1993 for a more comprehensive discussion). We will return to the last item, the possible confusion with primary fluctuations, in Section 3.

### 2.2.1 Instrumental noise

The instrumental noise is mainly due to the electronics of the receivers. It is usually assumed to be independent from pixel to pixel. This corresponds to a spatial power spectrum equivalent to that of white noise — see Tegmark & Efstathiou (1995), hereafter "TE95", for a detailed discussion. For a space-borne mission, the noise level currently achievable is of the order of $5\,(\theta/10')^{-1}\,\mu$K at 200 GHz (Mandolesi 1995).

### 2.2.2 Diffuse foreground emission

Atmospheric emission due to water vapour in the atmosphere will be at a level of about $50 - 100\,\mu$K (Fischer & Lange 1993) at a frequency of 250 GHz and below. At higher frequencies, it increases rapidly. This will make measurements of the kinematic SZ effect from the ground very difficult.

Outside the Earth's atmosphere, galactic synchrotron radiation and bremsstrahlung are the main sources of diffuse foreground emission at low frequencies (below the peak in the CMB spectrum), while at high frequencies, infrared emission from galactic dust and (young) galaxies is dominant. As we will discuss further on, the most favourable frequency for observing the kinematic SZ effect is somewhat above the peak in the CMB spectrum, at approximately 220 GHz. Here only the latter two effects potentially pose problems.

Our best information about the small-scale properties of the galactic dust comes from the 100, 60 and 25 $\mu$ IRAS maps. An extrapolation from these very high frequencies involves considerable uncertainties. However, the measurements of the kinematic SZ effect will only be sensitive to variation of the galactic dust emission on the rather small angular scales of ten arcminutes and below. Recent analyses of the IRAS maps (Low & Cutri 1994, TE95) indicate a spatial power spectrum for the galactic dust emission for which the power per decade in scale falls off rapidly towards smaller scales. A rough estimate of the expected dust contamination on 10' scales around 200 GHz is $10\,\mu$K in fairly clean parts of the sky (TE95). Since this contamination can be filtered out to some extent by using multi-frequency data, it is thus likely that dust will cause slightly less of a problem than instrumental noise.

The far infrared background emission due to the combined emission from (young) galaxies is hard to predict, as little is known about the properties of galaxies in the relevant wavelength range. Furthermore the evolution of number density, luminosity and spectrum of these objects towards higher redshift is highly uncertain. Theoretical estimates vary between a few and a few tens of $\mu$K, comparable to or slightly above the expected instrumental noise (Guiderdoni et al. 1995).

### 2.2.3 Radio sources

Discrete radio sources are usually one of the major difficulties for measurements of the Sunyaev-Zeldovich effect from the ground. However, these measurements are generally carried out at lower frequencies. The spectrum of the CMB rises rapidly ($\nu I_\nu \gtrsim \nu^3$ below of the peak of the spectrum at 160 GHz), much faster than typical radio source spectra. Radio source contamination will therefore be much less of a problem at the optimal frequency of 218 GHz.

### 2.2.4 The thermal SZ effect

The thermal SZ effect is generally considerably larger than the kinematic SZ effect. The exact factor of course depends on the density-weighted electron temperature and the peculiar velocity of the cluster. In the Raleigh-Jeans limit, the corresponding ratio of the observed change in brightness temperature is approximately a factor $20\,(T_{\rm e}/10^8 K)(v_{\rm pec}/500\,{\rm km\,s^{-1}})^{-1}$. However, as discussed in Section 2.1, the two effects are distinguishable by their different spectra. In the case of the thermal SZ effect, the change in the observed brightness temperature is a Compton distortion with a decrease at low frequencies, an increase at high frequencies and a zero-crossing at $\sim 218\,$GHz. The kinematic SZ effect, in contrast, has the same sign for all frequencies.

### 2.2.5 Bulk motions within the cluster

The bulk motion *within* the cluster (including a possible rotation) of the hot gas is expected to be small if the cluster is fully relaxed and the gas has attained hydrostatic equilibrium. However, evidence is increasing that galaxy clusters spend only a fraction of their lifetime in such an equilibrium state due to frequent merger events (Navarro, Frenk & White 1995). Internal bulk motions of the gas could therefore be frequent and of considerable amplitude. While these motions and the corresponding Doppler effect will be less relevant for poorly resolved or unresolved objects, they might be an important source of noise when filtering techniques are applied for extended objects. We will give a short discussion of this effect in Section 4.3.

### 2.2.6 Primary fluctuations

Observations of the DMR experiment on board the COBE satellite have established the existence of temperature fluctuations of the CMB on scales of a few degrees and above and determined their amplitude to be of order $(\Delta T/T) \sim 10^{-5}$ (Smoot *et al.* 1992). In the last two years, this result has been confirmed by a number of ground- and balloon-based experiments sensitive to temperature fluctuations on angular scales from a few degrees down to a few arcminutes. The scatter between the different measurements is still about a factor of two, but the measured amplitudes are consistent with the assumption that the fluctuations are generated by primordial potential fluctuations extending over all scales. In standard cosmological scenarios assuming a flat (Harrison-Zeldovich) power spectrum of potential fluctuations, the amplitude of temperature fluctuations should rise towards smaller scales and reach a maximum around 10 arcminutes (a scale set by the horizon size at the time when radiation and matter-density were equal). At even smaller scales a dramatic drop should occur due to diffusive damping (White, Scott & Silk 1994 for a recent general review on the CMB and Hu, Sugiyama & Silk 1995 for a description of the physics responsible for the Doppler peaks). Unfortunately, the spectral distribution of these intensity fluctuations is exactly the same as that of the kinematic SZ effect, *i.e.*, they correspond to a blackbody spectrum whose temperature varies slightly with position in the sky. For scales larger than the diffusive damping scale, cosmic fluctuations will therefore be the dominant noise source.

## 2.3 Isolating the kinematic SZ effect

The presence of the numerous noise sources described above make a multi-wavelength measurement absolutely essential to get a handle on possible systematic errors. In principle, modeling of noise components with different spectral signature should allow a discrimination of as many components as there are observed frequencies. In practice, when faced with both detector noise and considerable uncertainties in such modeling, a more robust approach is to choose a suitable frequency and estimate the impact of the dominant sources of contamination at that frequency. The simplest strategy is probably the use of a channel centered around the cross-over frequency of the thermal SZ effect (218 GHz), so that negative and positive contributions from lower and higher frequencies effectively cancel (Rephaeli & Lahav 1991). This channel would still be in the favourable frequency range where the diffuse foreground contamination is small. As an additional benefit, the radio source contamination will be much smaller than at the lower frequencies generally used so far for SZ measurements from the ground (*cf.* Willbanks et al. 1994).

# 3 Optimal filtering to minimize confusion with "cosmic" fluctuations

## 3.1 Estimating the noise due to cosmic fluctuations

### 3.1.1 The estimate for the peculiar velocity and its error

To get an estimate $V_{\rm est}$ for the radial peculiar velocity of a cluster, we need some information about the gas distribution in the cluster and the corresponding optical depth (which today would be obtained from X-ray observations, but in the future might also come from observation of the thermal SZ effect). If the cluster is resolved, we can choose an appropriately normalized filter function $\psi(\mathbf{x})$, and choose our estimate to be

$$V_{\rm est} = \int_F \frac{\Delta T}{T}(\mathbf{x})_{\rm obs}\, \psi(\mathbf{x})\, d^2x. \tag{5}$$

The filter function should be normalized so that

$$v_{\rm pec} \equiv \int_F S(\mathbf{x})\, \psi(\mathbf{x})\ \mathrm{d}^2x \tag{6}$$

is the true density weighted peculiar velocity of the cluster — $S(\mathbf{x})$ again denotes the noise-free signal. As shown in the appendix, the mean squared error of the estimator is given by

$$\langle (V_{\rm est} - v_{\rm pec})^2 \rangle = \frac{1}{(2\pi)^2} \int_F \widehat{\psi}(\mathbf{k})^2\ P(\mathbf{k})\ d^2k, \tag{7}$$

if the CMB "noise" properties are translationally invariant and can be described by a power spectrum $P(\mathbf{k})$. Here and throughout, hats denote Fourier transforms.

### 3.1.2 The assumed cluster profile and a first estimate of the expected noise level

To demonstrate the most important effects, we will use axisymmetric $\beta$-models to model the distribution of the cluster gas;

$$n_{\rm e}(r) = n_{\rm e0} \left[1 + \left(\frac{r}{r_{\rm c}}\right)^2\right]^{-3\beta/2}, \tag{8}$$

where $n_{\rm e0}$ is the central electron density and $r_{\rm c}$ is the core radius (Cavaliere & Fusco-Femiano 1976, 1978). With $0.5 < \beta < 1$, these models are generally a good fit to the electron density inferred from X-ray observations (Jones & Forman 1984). The optical depth of the cluster is then given by

$$\tau(\mathbf{x}) = \int n_e\, \sigma_{\rm T}\ {\rm d}l = n_{\rm e0}\, \sigma_{\rm T}\, r_{\rm c}\, \sqrt{\pi}\, \frac{\Gamma(3\beta/2 - 1/2)}{\Gamma(3\beta/2)} \left[1 + \left(\frac{\Theta}{\Theta_{\rm c}}\right)^2\right]^{-\frac{3\beta-1}{2}} . \quad (9)$$

Typical central optical depths for luminous X-ray clusters are of order 0.002–0.02. We now calculate the expected signal-to-noise as a function of the core radius of the cluster using a "naive" filter function $\psi(\mathbf{x}) \propto \tau'(\mathbf{x})$ appropriately normalized to get a first estimate of the expected noise level. $\tau'(\mathbf{x})$ denotes the convolution of the cluster profile with the beam profile.

In Figure 1, the corresponding $1\sigma$ error is shown for three different cosmological models. The following points should be noted here:

- The overall noise level is of order $250\,\mu$K, corresponding to about $1500\,(\tau/0.02)^{-1}\,{\rm km\,s^{-1}}$ (rather independent of the core radius for large core radii). This is a factor 10-50 higher than the other noise sources discussed in Section 2, and even for this optimistic optical depth, the expected $1\sigma$ error in the peculiar velocity measurement is rather large compared to typical cluster velocities.
- Beam dilution sets in when the core radius of the cluster becomes smaller than the resolution of the instrument. The signal-to-noise then scales as $\Theta_{\rm c}^{-1}$.
- The overall noise level depends only very weakly on the cosmological model.

## 3.2 Optimal filtering

### 3.2.1 The general shape of the filter function

A glance at equation (7) shows that the expected signal-to-noise depends on the filter function chosen. If we know the noise power spectrum well, we can optimize the filter function by minimizing the variance of our estimator. This is a simple variational problem, whose solution is given in Appendix A. We find that the Fourier transform of the optimal filter function $\psi$ is given by

$$\widehat{\psi}(\mathbf{k}) = \left[\frac{1}{(2\pi)^2}\int \frac{|\widehat{\tau'}(\mathbf{k})|^2}{P(\mathbf{k})} d^2k\right]^{-1} \frac{\widehat{\tau'}(\mathbf{k})}{P(\mathbf{k})}, \quad (10)$$

where $\widehat{\tau'}$ is the Fourier transform of $\tau'$. The corresponding signal-to-noise is given by

$$\frac{\mathrm{S}}{\mathrm{N}} = \frac{v_{\mathrm{pec}}}{\sqrt{\langle (V_{\mathrm{est}} - v_{\mathrm{pec}})^2 \rangle}} = \sqrt{\frac{1}{(2\pi)^2} \int \frac{\widehat{\tau'}(\mathbf{k})^2}{P(\mathbf{k})} \mathrm{d}^2\mathbf{k}}. \tag{11}$$

For Gaussian noise, the interpretation is simple, as the error in the estimate will then also be Gaussian distributed. Figure 2 shows the profile of typical filter functions. It has alternating positive and negative "sidelobes" which effectively cancel the long-wavelength cosmic fluctuations. The cluster profile is shown for comparison (both in arbitrary units).

#### 3.2.2 Axisymmetric filter

To demonstrate the main effects of optimal filtering, we again use axisymmetric $\beta$-profiles to characterize the radial profile of the gas distribution. Figures 3a-c show the effect of optimal filtering on the expected $1\sigma$ error of the estimate for the peculiar velocity as a function of the core radius of the cluster. The overall noise level is reduced by approximately a factor of two on all scales. This is due to the broad-band character of the cosmic fluctuations, with a substantial contribution from fluctuations with wavelengths much larger than the characteristic scale of the cluster — wavelengths which can efficiently be filtered out. However, the optimal filtering technique becomes even more efficient if the diffusive damping scale, below which the cosmic fluctuations are essentially zero, is above the characteristic scale of the cluster. The noise level can then be reduced to the pixel noise of the instrument if the instrumental resolution is sufficient. Figures 3a-c show the influence of the instrumental resolution, the pixel noise and the details of the cosmological model on the signal-to-noise ratio. The achieved signal-to-noise depends very sensitively on the size of the core radius of the cluster relative to the resolution of the instrument and the diffusive damping scale.

#### 3.2.3 Non-axisymmetric filter

If a high quality X-ray map of the cluster is available, we can do better than fitting a simple $\beta$-profile to the surface brightness distribution and constructing an axisymmetric filter function. To demonstrate this, we have used a surface brightness map of Coma (kindly provided by Hans Böhringer) and used the full two-dimensional spatial information to construct a non-axisymmetric spatial filter function. In Figure 4, the results are compared to axisymmetric filtering for different beam sizes and different values of the pixel noise. The cosmic noise is efficiently filtered out even for core radii comparable to or larger than the scale of the Doppler peaks. The achievable signal-to-noise now depends on the level of instrumental noise for all scales (as well as on the resolution of the instrument). However, as discussed in the next section, a spatial filter function which is optimized using such detailed information about the cluster gas distribution is more sensitive to errors in the assumed distribution.

#### 3.2.4 Possible complications

To use a non-axisymmetric optimal filter function, we need good knowledge of the spatial distribution of the line-of-sight integral of the radial velocity times the density $\int v_{\rm pec}(\mathbf{x})\, n(\mathbf{x}) {\rm d}l$. From the surface brightness distribution of a high-quality X-ray map we will get an estimate of the optical depth distribution. However, when constructing a spatial filter function, we will have to neglect the errors due to the unknown internal bulk motions, a possible clumping of the gas, the noise in the X-ray map and so on. To get at least a rough feeling for the effect of such errors in the assumed distribution of the signal, we have used clusters which were simulated with a Smoothed Particle Hydrodynamics code (kindly provided by Matthias Steinmetz) and were assumed to move with a peculiar velocity of $1000\,{\rm km\,s^{-1}}$. The simulations are similar to those described in Navarro, Frenk & White (1995) and show internal bulk motions of order $500\,{\rm km\,s^{-1}}$. We found that the estimated peculiar velocities were systematically smaller than the true optical depth-weighted mean values. The mean value of $v_{\rm est}/v_{\rm true}$ was 0.89 with a standard deviation of 0.18 for a sample of 12 cluster seen from three different directions. This should give an indication of the size of the effect of errors in the assumed signal distribution if non-axisymmetric filtering is applied. For axisymmetric filter functions, the errors were considerably smaller.

# 4 Determining cluster peculiar velocities

## 4.1 Using X-ray data to estimate the expected errors

The key parameters for a measurement of the kinematic SZ effect are the optical depth in the core, the core radius and the peculiar velocity of the cluster. In Appendix B, the luminosity-temperature relationship of galaxy clusters is used to derive the typical scaling of optical depth and core radius with X-ray luminosity, $\tau \sim L_{\rm X}^{5/11}$ and $r_c \sim L_{\rm X}^{-1/11}$. These scaling laws can be used to estimate optical depth and core radius relative to those of a reference cluster from the ratio of the X-ray luminosities and redshifts. In Figure 5, such estimates are shown for the XBACs cluster sample (X-ray brightest Abell-type clusters exceeding a flux limit of $5.5 \times 10^{-12}\,{\rm erg\,cm^{-2}\,s^{-1}}$ in the ROSAT band; Ebeling 1993, Ebeling et al. 1995). With the exception of a few well-studied clusters, the core radius and the central optical depth of a cluster are poorly determined quantities. We have therefore chosen Coma as a reference cluster for which we assume an optical depth $\tau = 0.01$ and a core radius of 10 arcminutes (Briel et al. 1992; see Herbig, Readhead & Lawrence 1992 for a measurement of the thermal SZ effect). However, one should note that these estimates will only be correct in a statistical sense, as there is a rather large spread in the luminosity-temperature relation of X-ray clusters reflecting the fact that there is no universal gas mass distribution for a cluster of given luminosity and redshift. Especially, cooling flow clusters will deviate from these scaling laws and will generally have smaller and denser cores (Edge & Stewart 1991; Fabian et al. 1994). Using Coma as a reference cluster should therefore yield conservative error estimates.

As expected, the angular extent of the core radii are generally smaller than for Coma, mainly due to the larger distances. Typical values lie between 1 and 3 arcminutes, corresponding to a typical redshift around $z = 0.1$. The optical depth lies between one fifth and twice that of Coma. We can now use the derived scaling laws to compute the expected $1\sigma$ error of a measurement of the kinematic SZ effect for the same cluster sample as above. In Figure 6, these are shown for our standard cosmological model ($\Omega_{\rm tot} = 1, \Omega_b = 0.06$) and two different beam sizes (4 and 8 arcminutes). Here we have assumed that an X-ray map will exist of sufficient quality to fit a $\beta$-model to the X-ray surface brightness, so that we can apply the filtering technique with an axisymmetric filter. As discussed above, there is a characteristic angular scale, above which confusion with primary fluctuations becomes important and which is about 5 arcminutes for the special cosmological model chosen here. The $1\sigma$ errors drop considerably for clusters with smaller core radii if the beam size is sufficiently small to "avoid the Doppler peaks". Even though this is difficult, it might be possible for a future space-borne mission. For a diffraction limited instrument with aperture $D$ observing at frequency $\nu$, the maximum resolution is approximately $5' \,(D/1\,{\rm m})^{-1}(\nu/200\,{\rm GHz})^{-1}$.

## 4.2 Individual clusters

For standard cosmological scenarios, it seems impossible to obtain more than upper limits of order $1000\,\mathrm{km\,s^{-1}}$ for the peculiar velocity of clusters with core radii larger than 3-7 arcminutes, even if the gas distribution is known well enough to apply non-axisymmetric filtering. This probably precludes a useful determination of the peculiar velocity of a nearby extended X-ray luminous cluster like Coma. However, chances are better for nearby cooling-flow clusters which have smaller core radii and a larger central optical depth (Edge & Stewart 1991). Generally, prospects are best for the most X-ray luminous clusters at somewhat higher redshifts, with core radii around 3 arcminutes. Figure 5 shows that there are approximately 30 such suitable clusters with a core radius around 3 arcminutes and $\tau \gtrsim 0.01$. The minimum $1\sigma$ error should here be around $\sim 300-700\,\mathrm{km\,s^{-1}}$. This might just be sufficient if the peculiar velocities are as high as indicated by recent peculiar velocity studies of Abell clusters out to a redshift of $15000\,\mathrm{km\,s^{-1}}$ (Lauer & Postman 1994; Postman & Lauer 1995; Colless 1995). For a diffraction-limited instrument with an aperture of about 1 m, the measurement of the kinematic SZ effect for clusters with smaller core radii will suffer considerably from beam dilution effects. The minimum measurable peculiar velocity therefore increases again towards very small core radii.

Figure 7 shows the distribution of $1\sigma$ errors of the cluster sample for different cosmological models and a beam size of 4 arcminutes. The errors generally increase if $\Omega_{\mathrm{tot}}$ is lowered and decrease if $\Omega_{\mathrm{bar}}$ is lowered.

## 4.3 Statistical measurement of bulk velocities on large scales

An obvious way to improve the signal-to-noise ratio is averaging the signal of ensembles of clusters. To get a feeling for what could be achieved, we have calculated the expected error for a maximum likelihood solution of the bulk flow velocity with respect to the microwave background, again for the XBACs cluster sample (Ebeling 1993). Neglecting small-scale peculiar velocities, the maximum likelihood solution for the bulk flow velocity $\mathbf{U}$ of a sample of a clusters at positions $\mathbf{r}_q$ with measured radial peculiar velocities $v_q$ is given by

$$U_i = A_{ij}^{-1} \sum_q \frac{\hat{r}_{q,j}\, v_q}{\sigma_q^2}, \qquad\qquad A_{ij} = \sum_q \frac{\hat{r}_{q,i}\, \hat{r}_{q,j}}{\sigma_q^2}, \tag{12}$$

where $\hat{\mathbf{r}} = \mathbf{r}/|\mathbf{r}|$ and the $\sigma_q$ are the rms peculiar velocity errors (Kaiser 1988). The covariance matrix of the maximum likelihood estimates is

$$\langle U_i U_j \rangle = A_{ij}^{-1}. \tag{13}$$

Table 1 shows the rms errors for the x, y and z components of $\mathbf{U}$ in galactic coordinates for a set of cosmological models and two beam sizes. The number of clusters in the sample is 244 and the mean redshift of the sample is $z = 0.088$. The values indicate that a reliable measurement of bulk velocities should be possible at redshifts, $cz \sim 10000 - 50000\,\mathrm{km\,s^{-1}}$ corresponding to scales of $100 - 500h^{-1}$ Mpc.

## 5 Conclusions

In view of the severe obstacle of atmospheric noise facing ground-based instruments, a space-borne mission seems to be the most promising method to measure the kinematic SZ effect. As pointed out previously, mapping the sky at a frequency of 218 GHZ, where the thermal SZ effect vanishes, should be the optimal strategy. The dominant noise source will here be confusion with primary fluctuations. The expected noise level is of order or larger than the expected signal, and it is therefore essential to use the knowledge of the CMB "noise" properties and the gas distributions of the individual clusters (which can be obtained by the mission itself and from X-ray observations, respectively). This knowledge makes it possible to analyze the CMB maps with a spatial filter optimized for individual clusters. An improvement in signal-to-noise by a factor of two is easily achievable, and even a factor of 10 is possible if the gas mass distribution is well known from a high-quality X-ray map. The final signal-to-noise ratio depends crucially on the cosmological model and the angular resolution of the instrument, and it is not currently clear whether a meaningful peculiar velocity measurement for individual clusters will be possible. Prime candidates are X-ray luminous clusters at intermediate redshift with core radii just below the Doppler peak scale. For a favourable but still rather standard cosmological scenario (standard CDM with low baryon fraction) and a good angular resolution (4' FWHM), the peculiar velocity of as many as 30 individual clusters might be determined accurately. Even if this is impractical, it should still be possible to determine the bulk motion of an ensemble of 200 X-ray luminous clusters at redshifts $\gtrsim 10000\,\mathrm{km\,s^{-1}}$ with an accuracy of order $200\,\mathrm{km\,s^{-1}}$.

We are grateful to Hans Böhringer, Harald Ebeling, Mathias Steinmetz and Naoshi Sugiyama for kindly providing X-ray data, SPH simulations and CMB power spectra, respectively. We also thank Hans Böhringer, Rashid Sunyaev and especially Simon White for helpful comments on the manuscript. This work was supported in part by "Sonderforschungsbereich 375-95 für Astro-Teilchenphysik der Deutschen Forschungsgemeinschaft" and by European Union contract CHRX-CT93-0120.

## Appendix A: Optimal filtering

In this Appendix, we derive the optimal filtering scheme which was used in the main part of the paper. Suppose we want to extract the amplitude $A$ of a signal with a known spatial distribution $\tau(\mathbf{x})$ from a measured signal $D(\mathbf{x})$ which is contaminated by noise $N(\mathbf{x})$. Writing

$$D(\mathbf{x}) = A\,\tau(\mathbf{x}) + N(\mathbf{x}), \tag{14}$$

the application in this paper corresponds to $\tau(\mathbf{x})$ being the known optical depth profile of some cluster, $A$ being proportional to the radial velocity $v_{\rm pec}$, and the random field $N$ being the nuisance contribution from the CMB and pixel noise. Let us make no other assumptions about $N$ than that it has zero mean and that its statistical properties are independent of position. Letting hats denote Fourier transforms, this implies that

$$\begin{aligned} \langle N(\mathbf{x})\rangle &= 0, & (15)\\ \langle \widehat{N}(\mathbf{k})^*\widehat{N}(\mathbf{k})\rangle &= (2\pi)^2\delta(\mathbf{k}'-\mathbf{k})P(\mathbf{k}), & (16) \end{aligned}$$

for some power spectrum $P(\mathbf{k})$. Note that we are not making any assumptions about $N$ being Gaussian. The most general estimator of $A$ that is linear in the data can clearly be written as

$$A_{\rm est} \equiv \int \psi(\mathbf{x})D(\mathbf{x})d^2x \tag{17}$$

for some weight function $\psi$. Using equations (15) and (16), we readily obtain

$$\begin{aligned} b &\equiv \langle A_{\rm est} - A\rangle = \int \psi(\mathbf{x})\tau(\mathbf{x})d^2x \,-\, 1, & (18)\\ \sigma^2 &\equiv \langle (A_{\rm est}-A)^2\rangle = b^2 + \frac{1}{(2\pi)^2}\int \left|\widehat{\psi}(\mathbf{k})\right|^2 P(k)d^2k, & (19) \end{aligned}$$

were $b$, the *bias*, gives the average error and $\sigma$, the *noise*, gives the r.m.s. deviation of the estimate from the true value. The goal of this Appendix is to find the best estimator $A_{\rm est}$, *i.e.*, the optimal choice of $\psi$. Now what do we mean by an estimator being *good*? We will use the following criteria:

- Require it to be unbiased ($b = 0$)
- Minimize the noise $\sigma$

The requirement of no bias clearly just fixes the normalization of $\psi$, whereas its shape will be determined by the second condition, the requirement that the average error be minimized. The problem of minimizing $\sigma^2$ subject to the constraint that $b = 0$ is readily solved by introducing a Lagrange multiplier $\lambda$ and finding the function $\psi$ that minimizes $L \equiv \sigma^2 + \lambda b$. Using Parseval's theorem to re-express $b$ in terms of $\widehat{\psi}$, we thus seek the function $\widehat{\psi}$ that minimizes

$$L = \frac{1}{(2\pi)^2}\int \widehat{\psi}(\mathbf{k})^*\left[\widehat{\psi}(\mathbf{k})P(\mathbf{k}) + \lambda\widehat{\tau}(\mathbf{k})\right] d^2k \; - \lambda, \tag{20}$$

Taking the functional derivative with respect to $\widehat{\psi}$ and setting the result equal to zero, we thus obtain the simple result $\widehat{\psi}(\mathbf{k}) \propto \widehat{\tau}(\mathbf{k})/P(\mathbf{k})$. This has a simple intuitive interpretation: the optimal filter $\psi$ is such that it is maximally sensitive to those wave numbers $\mathbf{k}$ where $\widehat{\tau}$ is large and $P$ is small. In other words, it focuses on those Fourier modes where the cluster profile stands out the most above the noisy background. Normalizing this so that $b = 0$, we get

$$\widehat{\psi}(\mathbf{k}) = \left[\frac{1}{(2\pi)^2}\int \frac{|\widehat{\tau}(\mathbf{k}')|^2}{P(\mathbf{k}')} d^2k'\right]^{-1} \frac{\widehat{\tau}(\mathbf{k})}{P(\mathbf{k})}, \tag{21}$$

our desired result. Substituting this back into equation (19), we find that the best signal-to-noise ratio $A/\sigma$ that can be attained is

$$\frac{A}{\sigma} = \left[\frac{1}{(2\pi)^2}\int \frac{|\widehat{\tau}(\mathbf{k})|^2}{P(\mathbf{k})} d^2k\right]^{1/2} A. \tag{22}$$

To apply these formulas in real-world situations, the effects of pixel noise and beam smoothing are simply absorbed into $P(k)$ as described in T95. The quantities of equation (14) all become convolved with the beam function $B$, and $N$ receives an additional contribution from pixel noise, leaving

$$D'(\mathbf{x}) = A\,\tau'(\mathbf{x}) + N'(\mathbf{x}). \tag{23}$$

Here $\tau'$ denotes $\tau$ convolved with the beam as before (*i.e.*, $\widehat{\tau}' = \widehat{B}\widehat{\tau}$), $D'$ denotes the noisy beam-smoothed signal that we actually observe, and as shown in T95, $N'$ has the power spectrum

$$P'(\mathbf{k}) = P_{\rm cmb}(\mathbf{k})|\widehat{B}(\mathbf{k})|^2 + \frac{4\pi\sigma_n^2}{n}. \tag{24}$$

Here $n/4\pi$ is the number of pixels per steradian, and $\sigma_n$ denotes the r.m.s. noise level per pixel. Substituting this into the solution we found above, we thus obtain the optimal filter function $\widehat{\psi}(\mathbf{k}) \propto \widehat{\tau}'(\mathbf{k})/P'(\mathbf{k})$, *i.e.*,

$$\widehat{\psi}(\mathbf{k}) = \left[\frac{1}{(2\pi)^2}\int \frac{|\widehat{\tau}(\mathbf{k}')|^2}{P_{\rm cmb}(\mathbf{k}') + \frac{4\pi\sigma_n^2}{n|\widehat{B}(\mathbf{k}')|^2}} d^2k'\right]^{-1} \frac{\widehat{\tau}(\mathbf{k})}{P_{\rm cmb}(\mathbf{k})\widehat{B}(\mathbf{k})^* + \frac{4\pi\sigma_n^2}{n\widehat{B}(\mathbf{k})}}. \quad (25)$$

Thus the best signal-to-noise ratio $A/\sigma$ that can be attained is

$$\frac{A}{\sigma} = \left[\frac{1}{(2\pi)^2}\int \frac{|\widehat{\tau}(\mathbf{k})|^2}{P_{\rm cmb}(\mathbf{k}) + \frac{4\pi\sigma_n^2}{n|\widehat{B}(\mathbf{k})|^2}} d^2k\right]^{1/2} A. \quad (26)$$

If the instrument has a Gaussian beam, then

$$\widehat{B}(\mathbf{k}) \approx \exp\left[-\theta^2\,(l+1)\,l\right], \quad (27)$$

where $\theta = \mathrm{FWHM}/2\sqrt{\ln 2} \approx 0.6\mathrm{FWHM}$ and FWHM denotes the full-width-half-max.

## Appendix B: Using scaling relations to estimate the optical depth and core radius of a cluster

It is generally accepted that the properties of clusters of galaxies obey certain scaling relations, even though the scatter is large (Edge & Stewart 1991). The scaling of the X-ray luminosity with the temperature of the emitting gas is the best established of these correlations, $L_{\rm X} \sim T^{\alpha}$ with $\alpha \approx 3$.

For bremsstrahlung emission from a cluster of temperature $T$ with core radius $r_c$ and core density $n_c$, the X-ray luminosity should scale as

$$L_X \sim n_c^2\, r_c^3\, T^{1/2}. \quad (28)$$

The temperature of the gas should be determined by the depth of the potential well of the gravitationally dominant dark matter halo in which the gas is believed to reside. If the gas traces the dark matter (as indicated by numerical simulations), the temperature of the gas should scale as $T_c \sim n_c\, r_c^2$, which translates into

$$L_X \sim T^{5/2}\, r_c^{-1}. \quad (29)$$

The observed scaling of luminosity with temperature is then obtained if the core radius mildly decreases with increasing luminosity. A nice theoretical explanation for such a scaling was first discussed by Kaiser (1991). Kaiser suggested that the gas in all cluster cores might have the same entropy as would be the case if the gas from which clusters form was preheated to a certain temperature. If the gas in cluster cores was then adiabatically compressed during the collapse of clusters, the core density should scale with temperature as $n_c \sim T_c^{3/2}$ for an adiabatic index of 5/3, which translates into $r_c \sim T_c^{-1/4}$ (Evrard & Henry 1993). This gives a scaling of luminosity $L \sim T^{11/4}$ and core radius

$$r_c \sim L^{-1/11}. \tag{30}$$

The optical depth should then scale as

$$\tau \sim n_c\, r_c \sim T\, r_c^{-1} \sim T^{5/9} \sim L^{5/11}. \tag{31}$$

Table 1: The estimated rms error of the x/y/z-component (galactic coordinates) of the bulk flow velocity of the XBACs cluster sample with respect to the cosmic microwave background for different cosmological models and two beam sizes.

| | | beam size | |
|---|---|---|---|
| | | 4' | 8' |
| | $\Omega_{\rm bar} = 0.01$ | 200/193/130 km s$^{-1}$ | 304/307/202 km s$^{-1}$ |
| $\Omega_{\rm tot} = 0.1$ | $\Omega_{\rm bar} = 0.03$ | 207/200/135 km s$^{-1}$ | 321/324/213 km s$^{-1}$ |
| | $\Omega_{\rm bar} = 0.06$ | 212/204/138 km s$^{-1}$ | 336/339/223 km s$^{-1}$ |
| | $\Omega_{\rm bar} = 0.01$ | 134/129/87 km s$^{-1}$ | 211/209/138 km s$^{-1}$ |
| $\Omega_{\rm tot} = 0.3$ | $\Omega_{\rm bar} = 0.03$ | 159/153/103 km s$^{-1}$ | 240/239/158 km s$^{-1}$ |
| | $\Omega_{\rm bar} = 0.06$ | 166/160/108 km s$^{-1}$ | 247/246/163 km s$^{-1}$ |
| | $\Omega_{\rm bar} = 0.01$ | 54/52/35 km s$^{-1}$ | 97/96/64 km s$^{-1}$ |
| $\Omega_{\rm tot} = 1.0$ | $\Omega_{\rm bar} = 0.03$ | 68/65/44 km s$^{-1}$ | 137/134/90 km s$^{-1}$ |
| | $\Omega_{\rm bar} = 0.06$ | 100/96/65 km s$^{-1}$ | 165/163/108 km s$^{-1}$ |

## Figure captions

Figure 1: The $1\sigma$ error for the determination of the peculiar velocity as a function of the core radius of the cluster using a "naive" filter function ($\psi(\mathbf{x}) \propto \tau'(\mathbf{x})$) for three different cosmological models. The inset shows the angular power spectrum $C_\ell$ of temperature fluctuations for the three models (plotted as usual as $\ell(\ell+1)C_\ell$ against multipole $\ell$). The solid curve is for a universe of critical density ($\Omega_{\rm tot} = 1$), a Hubble constant of $50\,\mathrm{km\,s^{-1}\,Mpc^{-1}}$ and a baryonic mass fraction of 6% ($\Omega_{\rm bar} = 0.06$). The dotted and dashed curves correspond to $\Omega_{\rm tot} = 0.1$. For the dashed curve, a non-standard reionization history is assumed (corresponding to an optical depth to Thomson scattering of order unity). The model for the gas distribution is an axisymmetric $\beta$-model with core radius $r_c$ ($\tau(r) \propto [1 + (r/r_c)^2]^{-3\beta/2}$, $\beta = 0.75$). The results are scaled to an optical depth of the cluster of $\tau = 0.02$ and scale with $\tau^{-1}$.

Figure 2:
a) Typical filter function for an axisymmetric cluster in arbitrary units. The cluster profile is shown for comparison. The angular scales are the same.
b) same as a) for a non-axisymmetric cluster. The cluster is shown at the top, the corresponding filter function at the bottom.

Figure 3:
a) The $1\sigma$ error in the determination of the peculiar velocity as a function of the core radius of the cluster using an axisymmetric "optimal" filter function ($\tilde{\psi}(\mathbf{k}) \propto \tilde{\tau}'(\mathbf{k})/P(\mathbf{k})$) for our fiducial standard cosmological scenario ($\Omega_{\rm tot} = 1$, $H_0 = 50\,{\rm km\,s^{-1}\,Mpc^{-1}}$, and $\Omega_{\rm bar} = 0.06$). The beam size varies as indicated in the plot. The pixel noise is fixed and corresponds to $7\,\mu$K in the 4' beam . The assumed gas distribution and central optical depth are the same as in Fig. 1. The dashed curve shows the results for the "naive filter" and a beam size of 8' for comparison.
b) Same as a) but for different values of the pixel noise. The pixel noise per beam varies as indicated in the plot. The assumed beam size is 4 arcminutes (FWHM).
c) Same as a) but for a different baryonic fraction and two different beam sizes. Thick curve are for a beam size of 8' and thin curves are for a beam size of 4'. the baryonic fraction $\Omega_{\rm bar}$ varies as indicated in the plot. The inset shows the angular power spectra of temperature fluctuations for the three cosmological models.

Figure 4:
a) The $1\sigma$ error for the determination of the peculiar velocity as a function of the core radius of the cluster. The solid lines are for a non-axisymmetric optimal filter function for our fiducial standard cosmological scenario ($\Omega_{\rm tot} = 1$, $H_0 = 50\,{\rm km\,s^{-1}\,Mpc^{-1}}$, and $\Omega_{\rm bar} = 0.06$). A scaled surface brightness map of Coma was used to construct the filter function. The beam size (FWHM) varies as indicated in the plot. The pixel pixel noise is fixed and corresponds to $7\,\mu$K in the 4' beam. The dashed curves show the results for an axisymmetric filter function for comparison (using the appropriately scaled best fitting $\beta$-model of Coma to construct the filter function).
b) Same as a) but for different values of the pixel noise. The pixel noise per beam varies as indicated in the plot. The assumed beam size is 4' (FWHM).

Figure 5: The optical depth and core radius of the XBACs cluster sample (Ebeling 1993) derived by using simple scaling laws, $\tau \sim L_X^{5/11}$, $r_c \sim L_X^{-1/11}$.

Figure 6: The expected $1\sigma$ errors of peculiar velocity measurements for the XBACs cluster sample (Ebeling 1993) for our fiducial standard cosmological model ($\Omega_{\rm tot} = 1$, $H_0 = 50\,{\rm km\,s^{-1}\,Mpc^{-1}}$, $\Omega_{\rm bar} = 0.06$) and a beam size of 8' and 4', respectively. The assumed optical depths and core radii are as in Fig 5. An axisymmetric optimal filter was applied using a $\beta$-model for the cluster gas distribution. The pixel noise is fixed and corresponds to $7\,\mu$K in the 4' beam.

Figure 7: The distribution of expected $1\sigma$ errors of peculiar velocity measurements for the XBACs cluster sample (Ebeling 1993) for different cosmological models. $\Omega_{\rm tot}$ and $\Omega_{\rm bar}$ vary as indicated in the plot. The assumed optical depth and core radius are as in Fig 5. An axisymmetric optimal filter was applied using a $\beta$-model for the cluster gas distribution. The pixel noise is fixed and corresponds to $7\,\mu$K in the 4' beam.

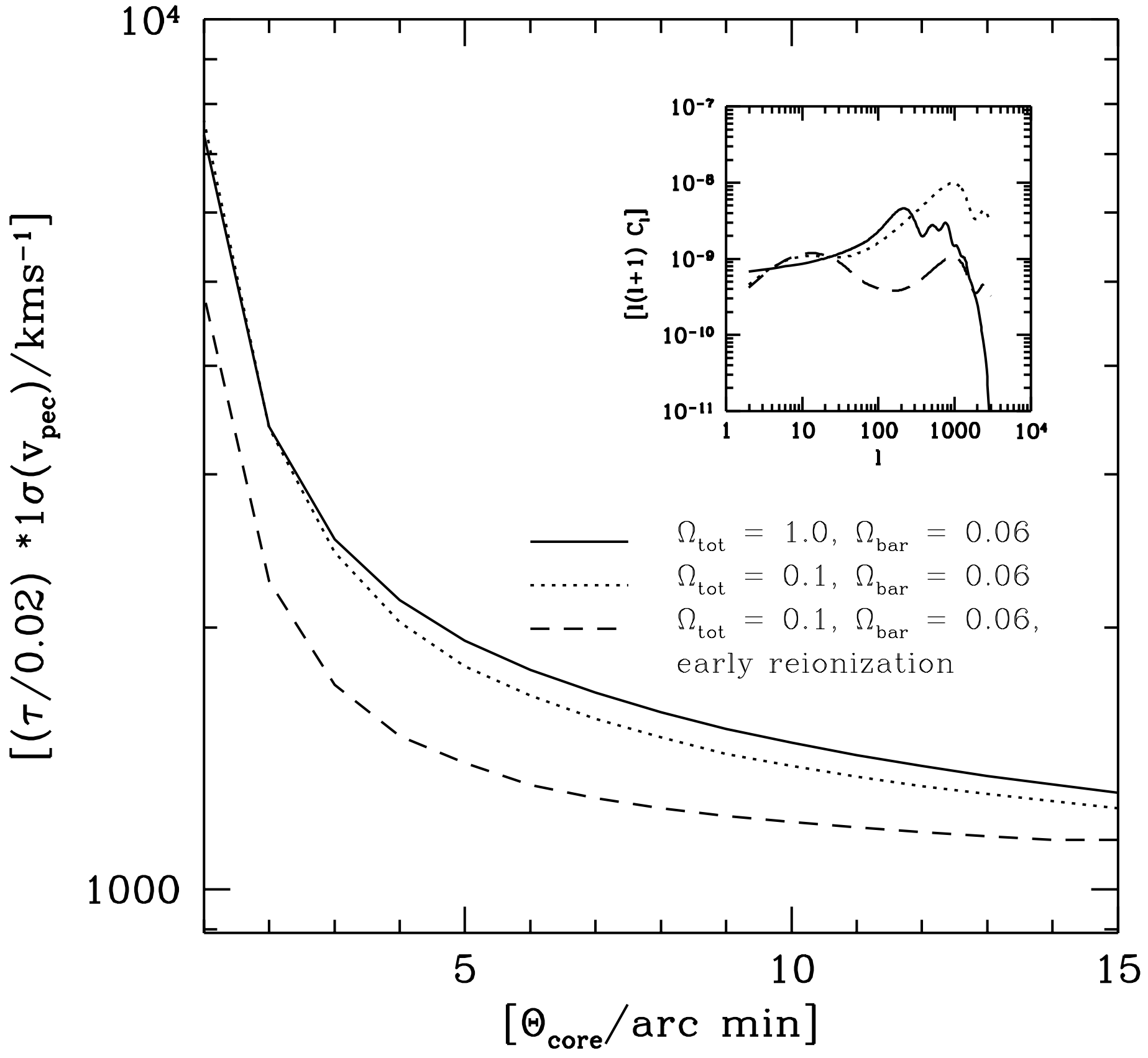

**Figure 1**

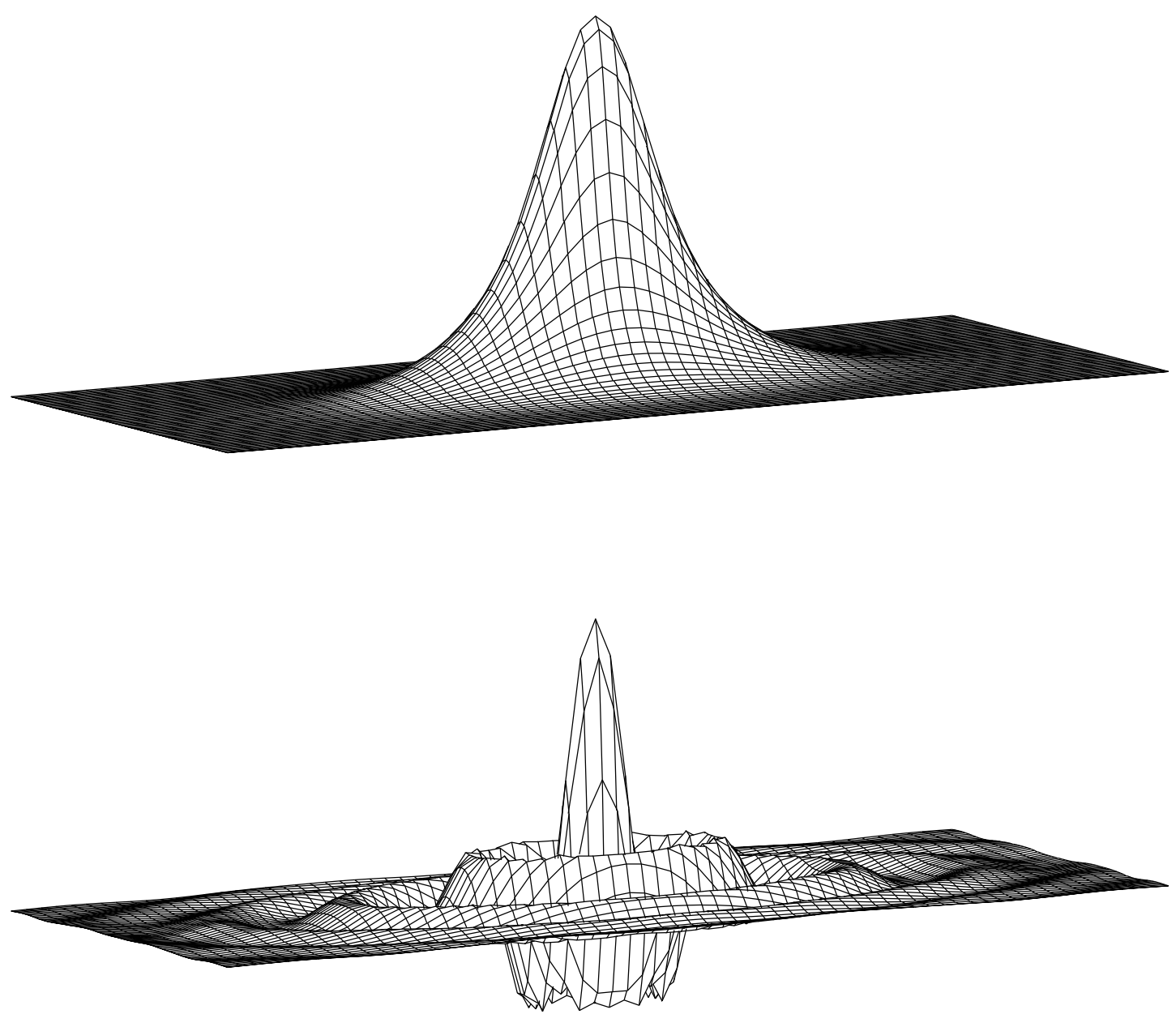

**Figure 2a**

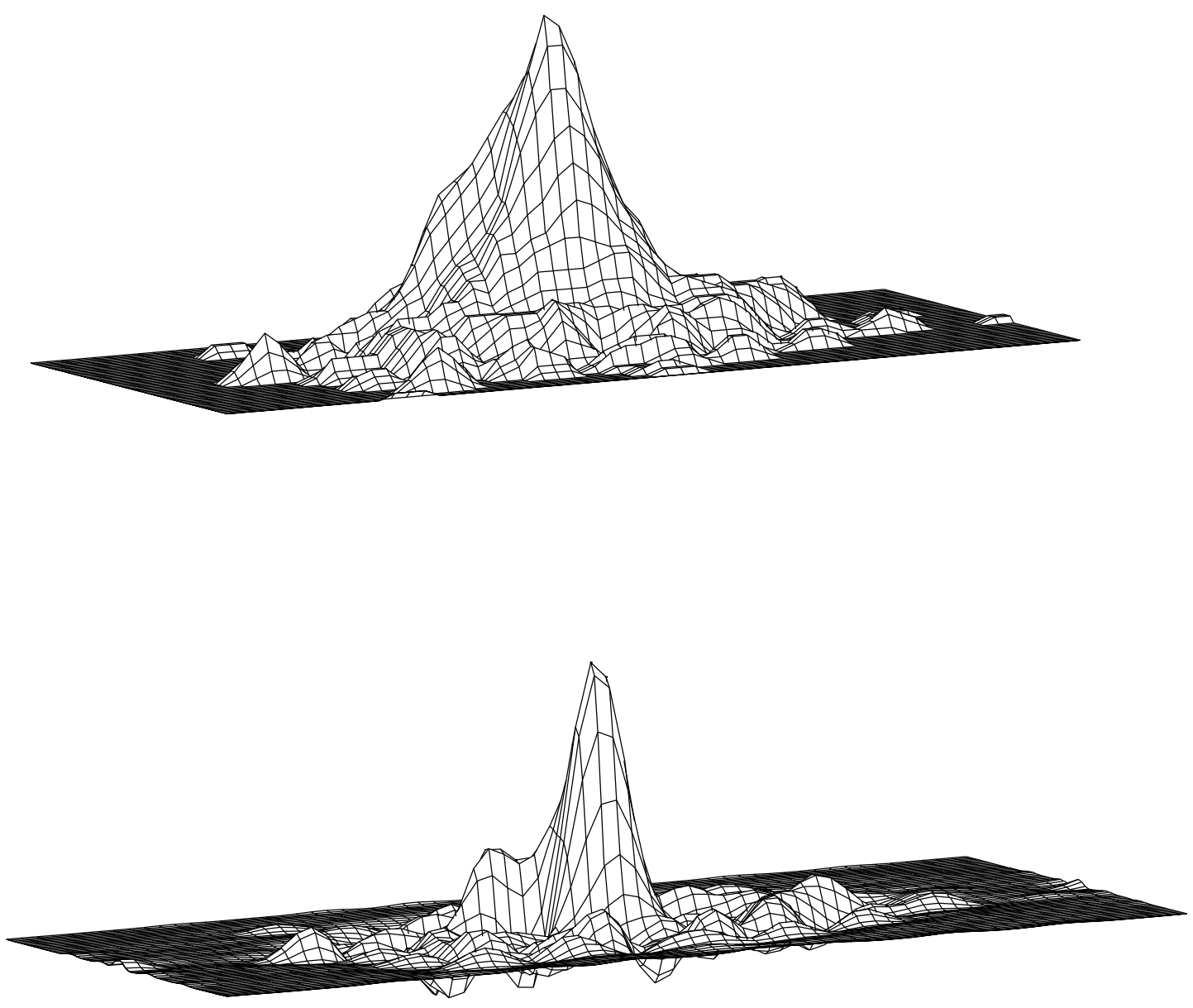

**Figure 2b**

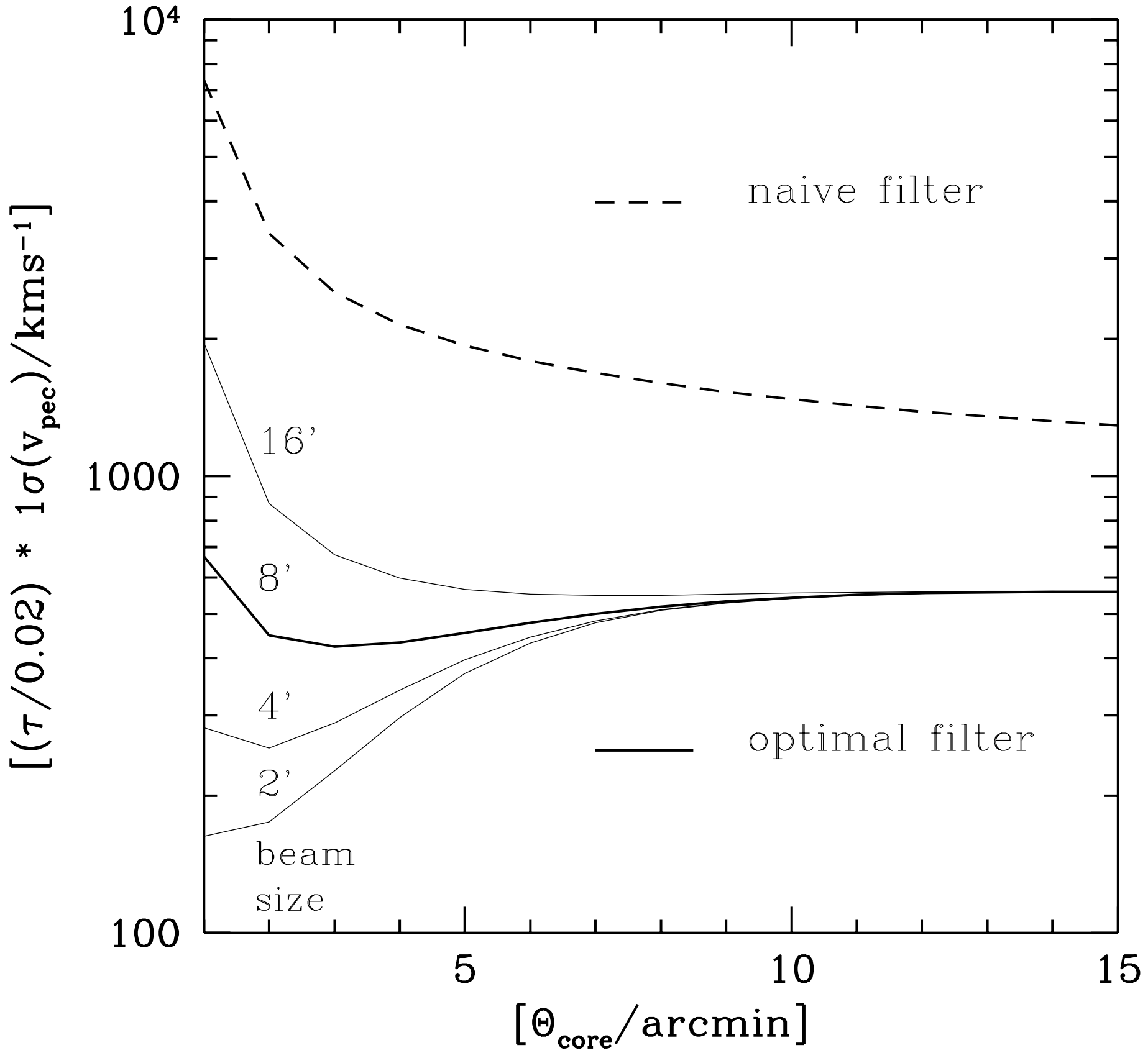

**Figure 3a**

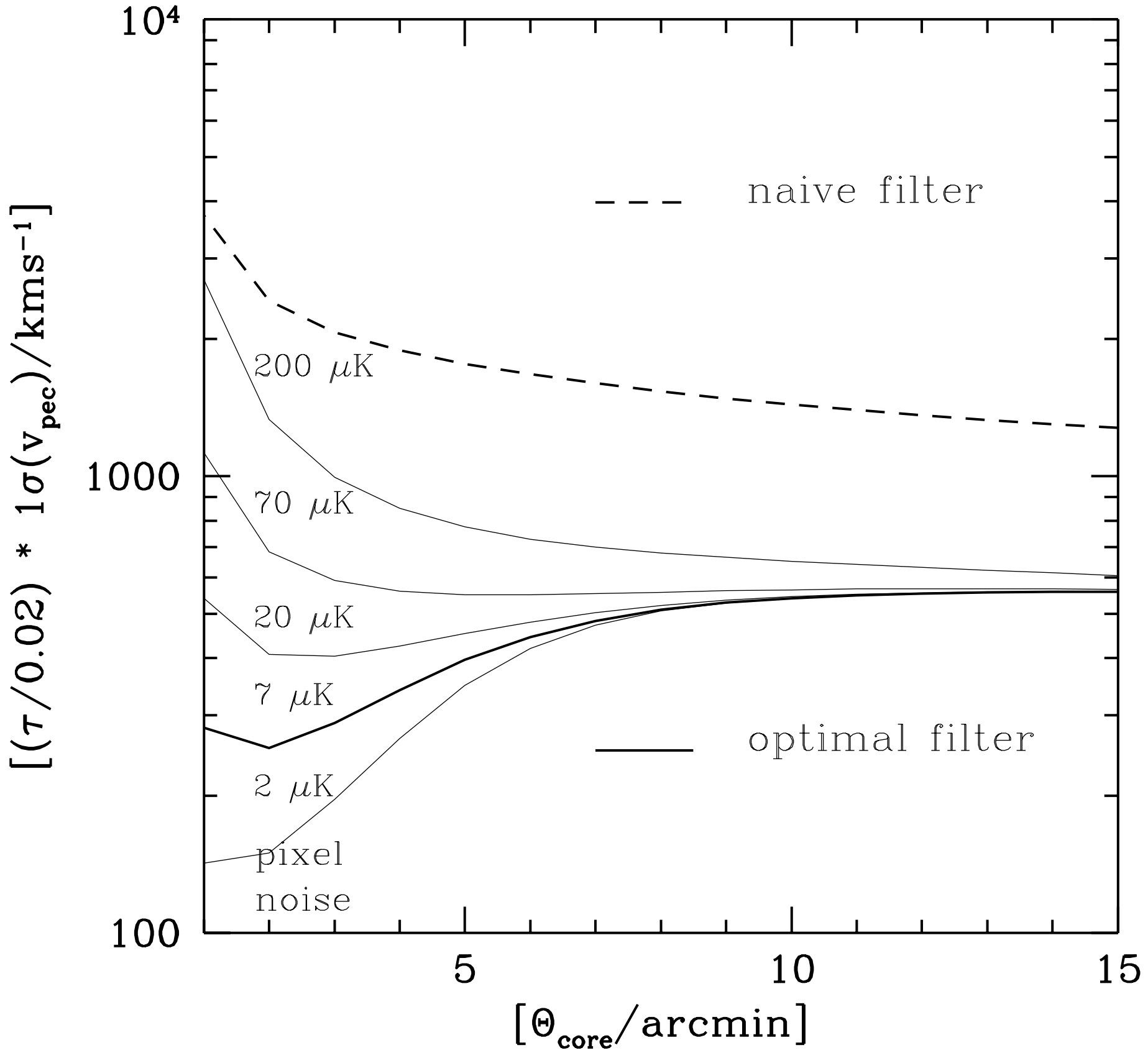

**Figure 3b**

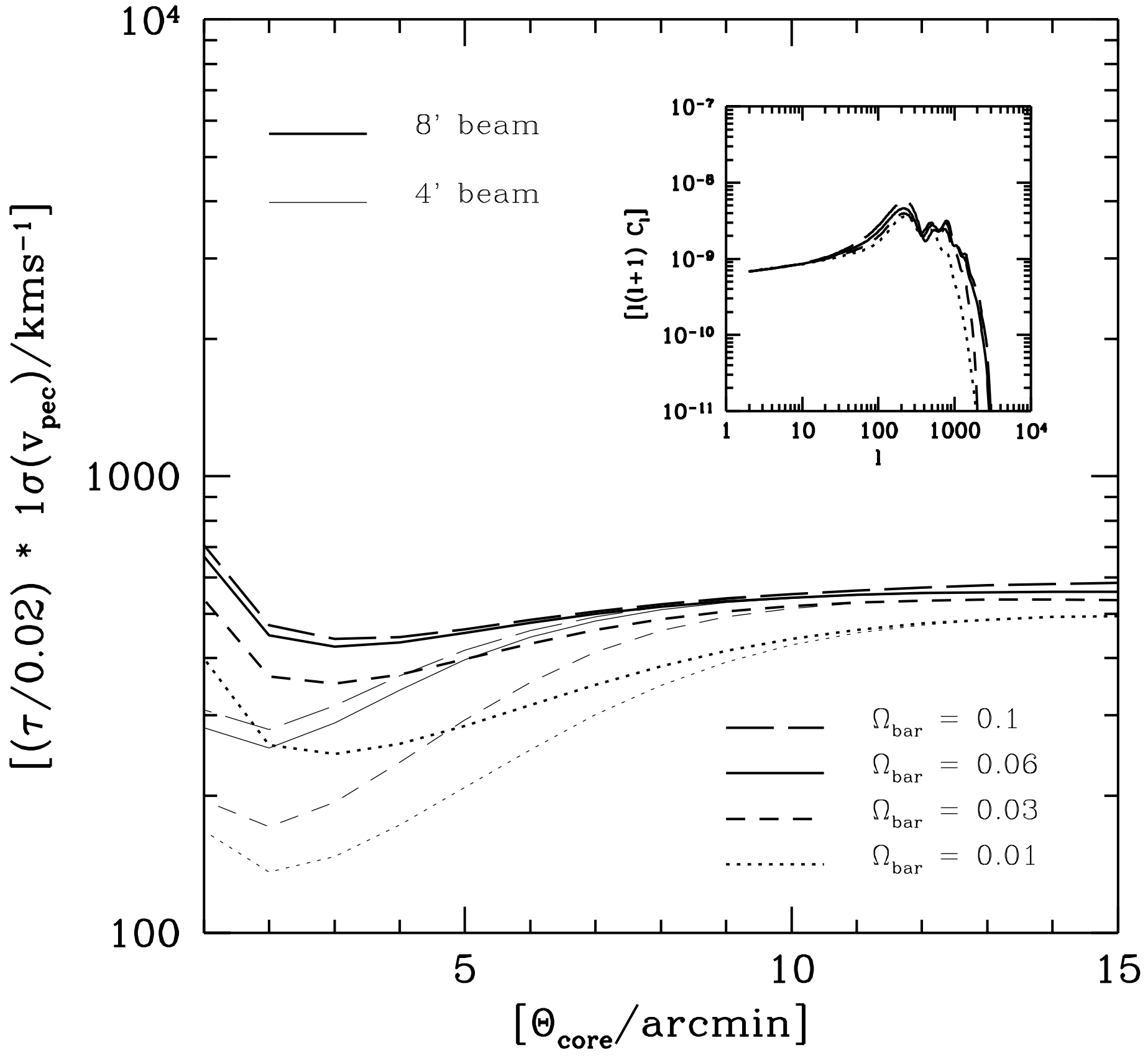

**Figure 3c**

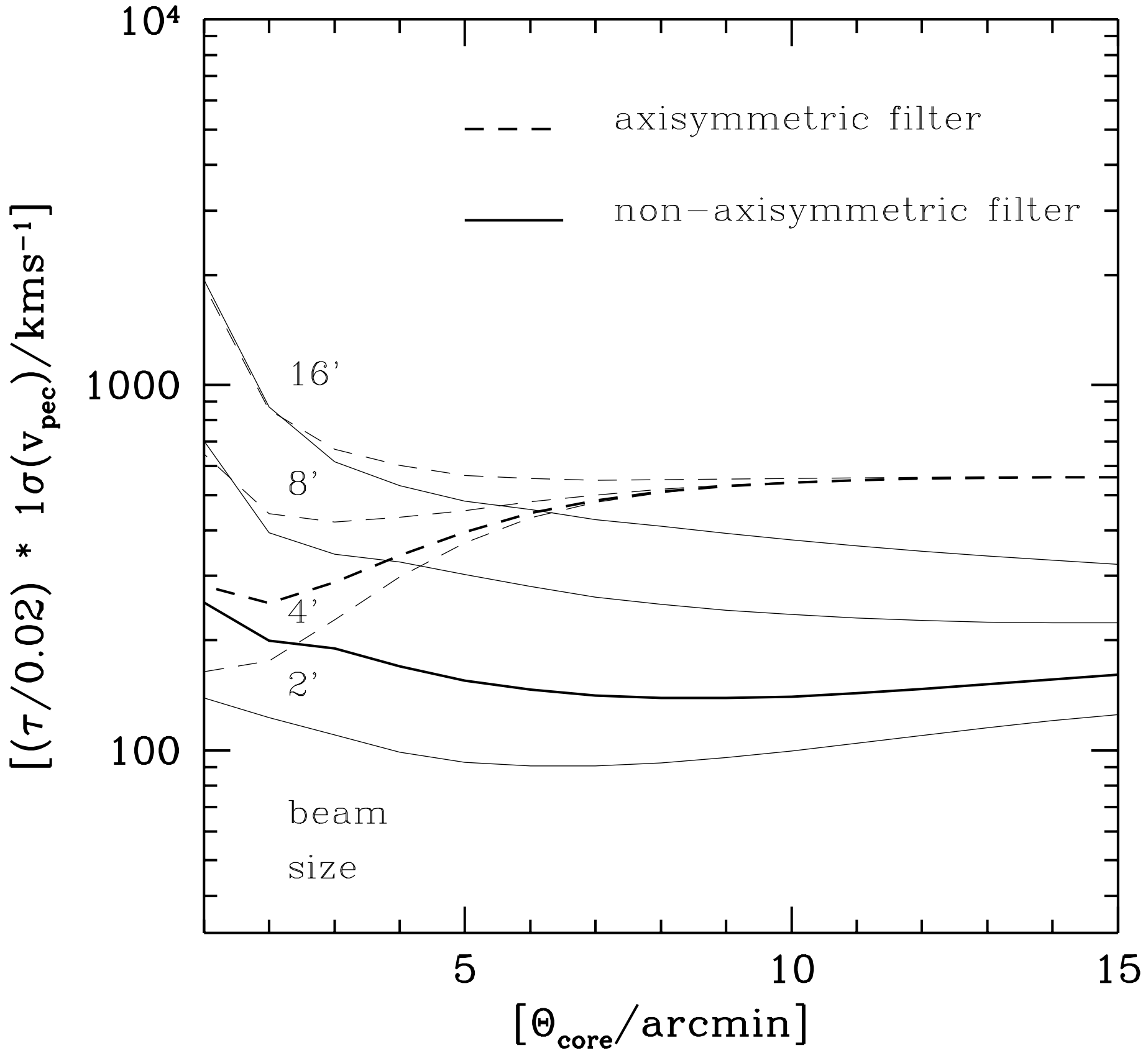

**Figure 4a**

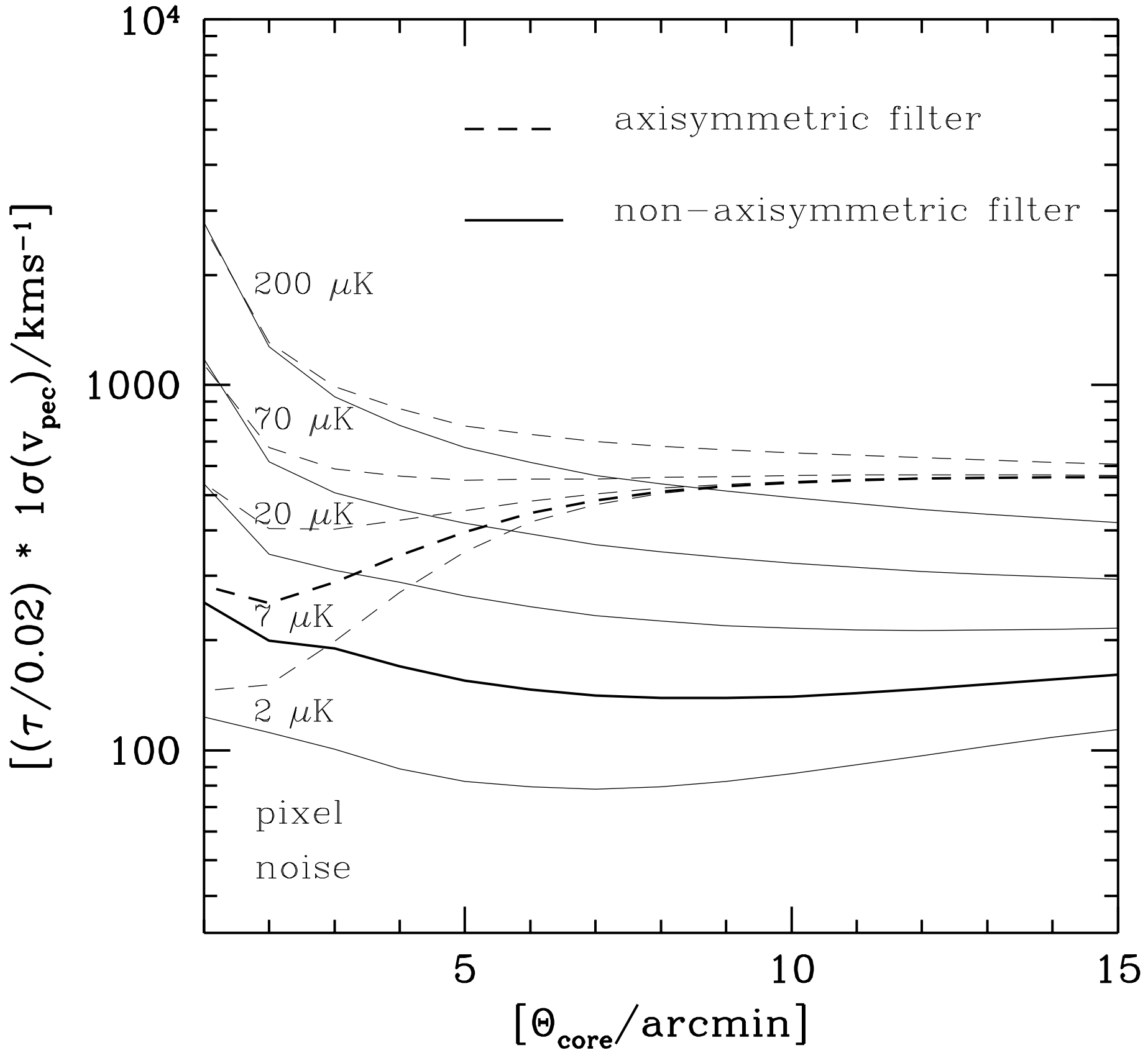

**Figure 4b**

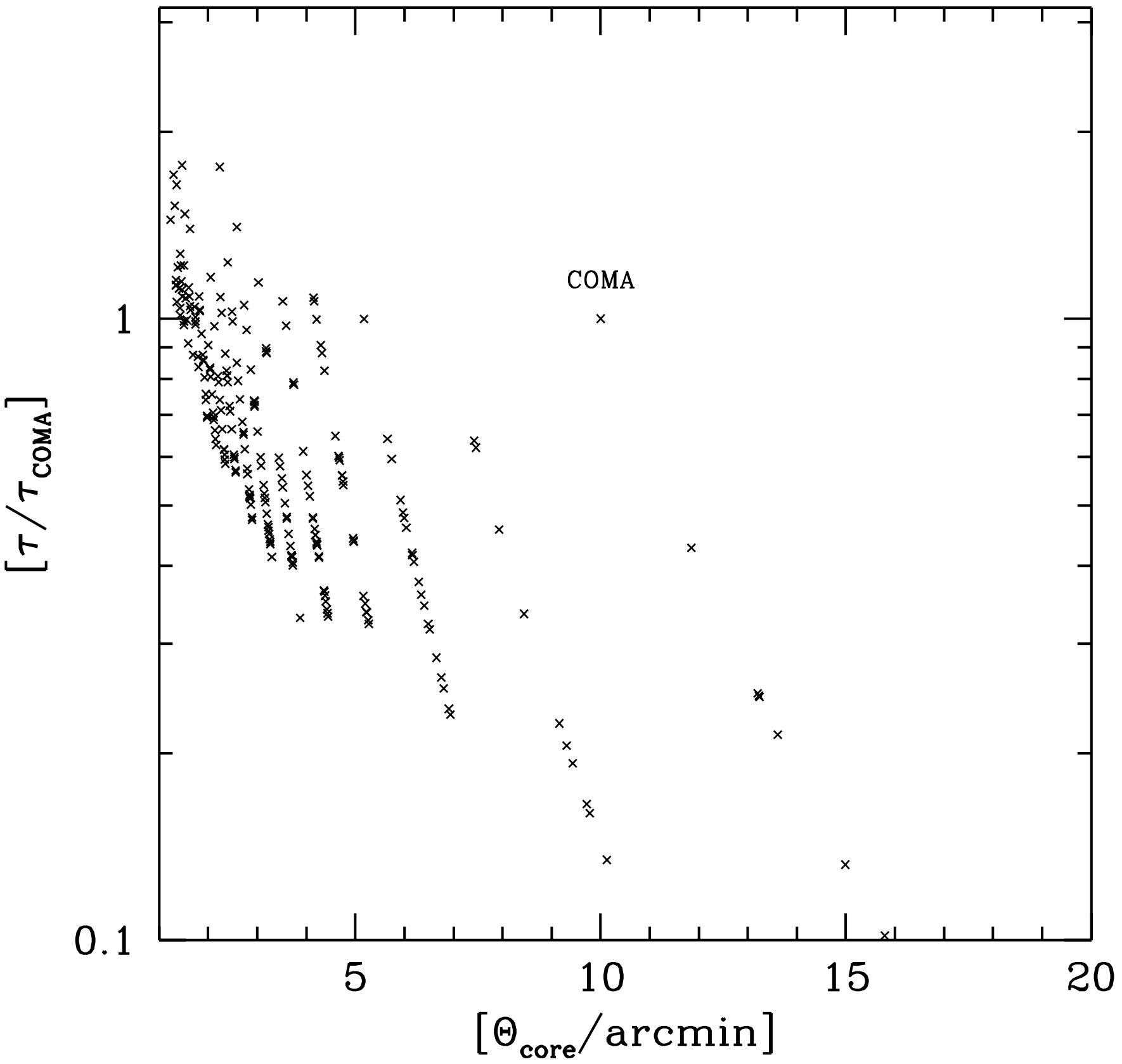

**Figure 5**

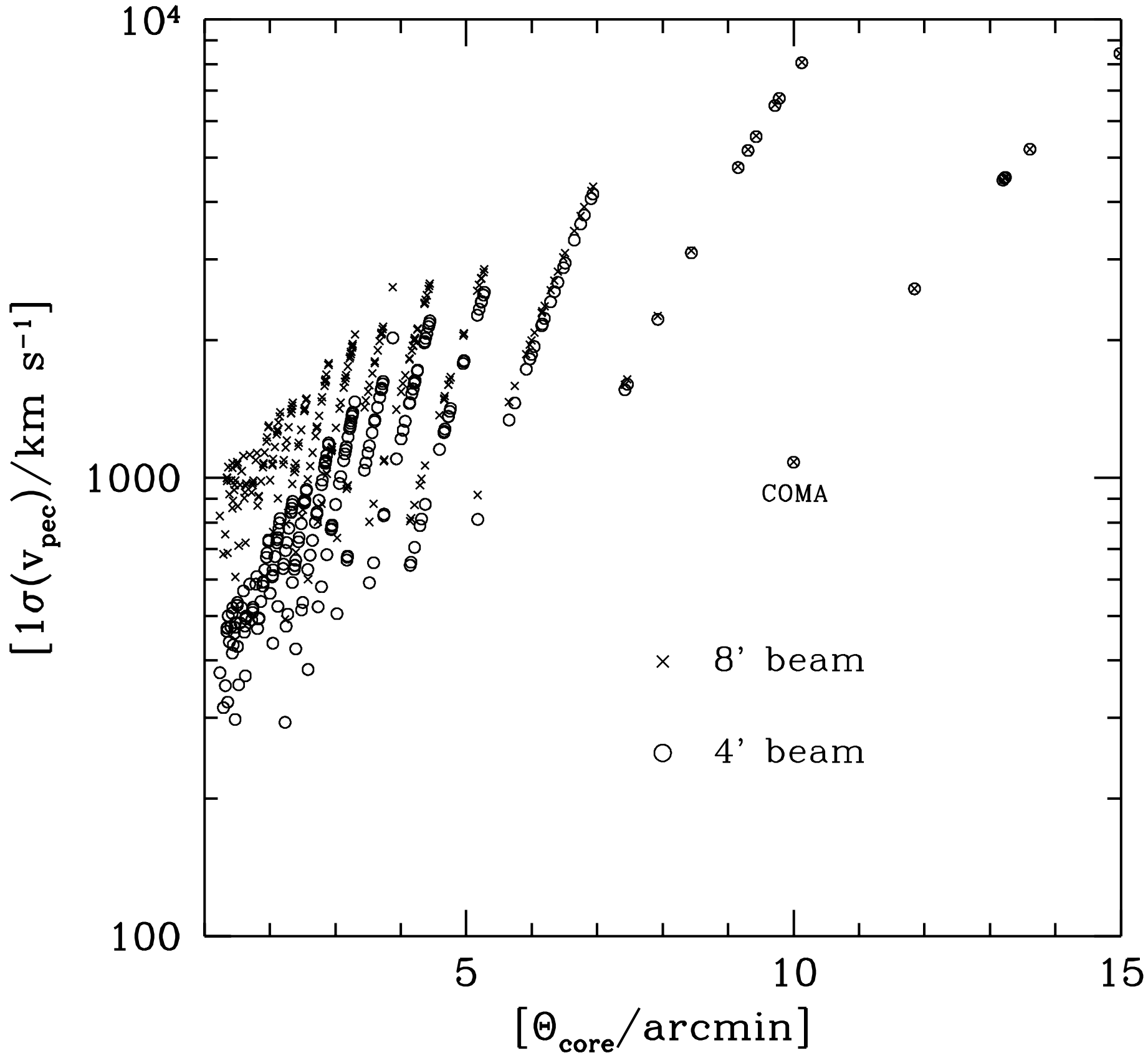

**Figure 6**

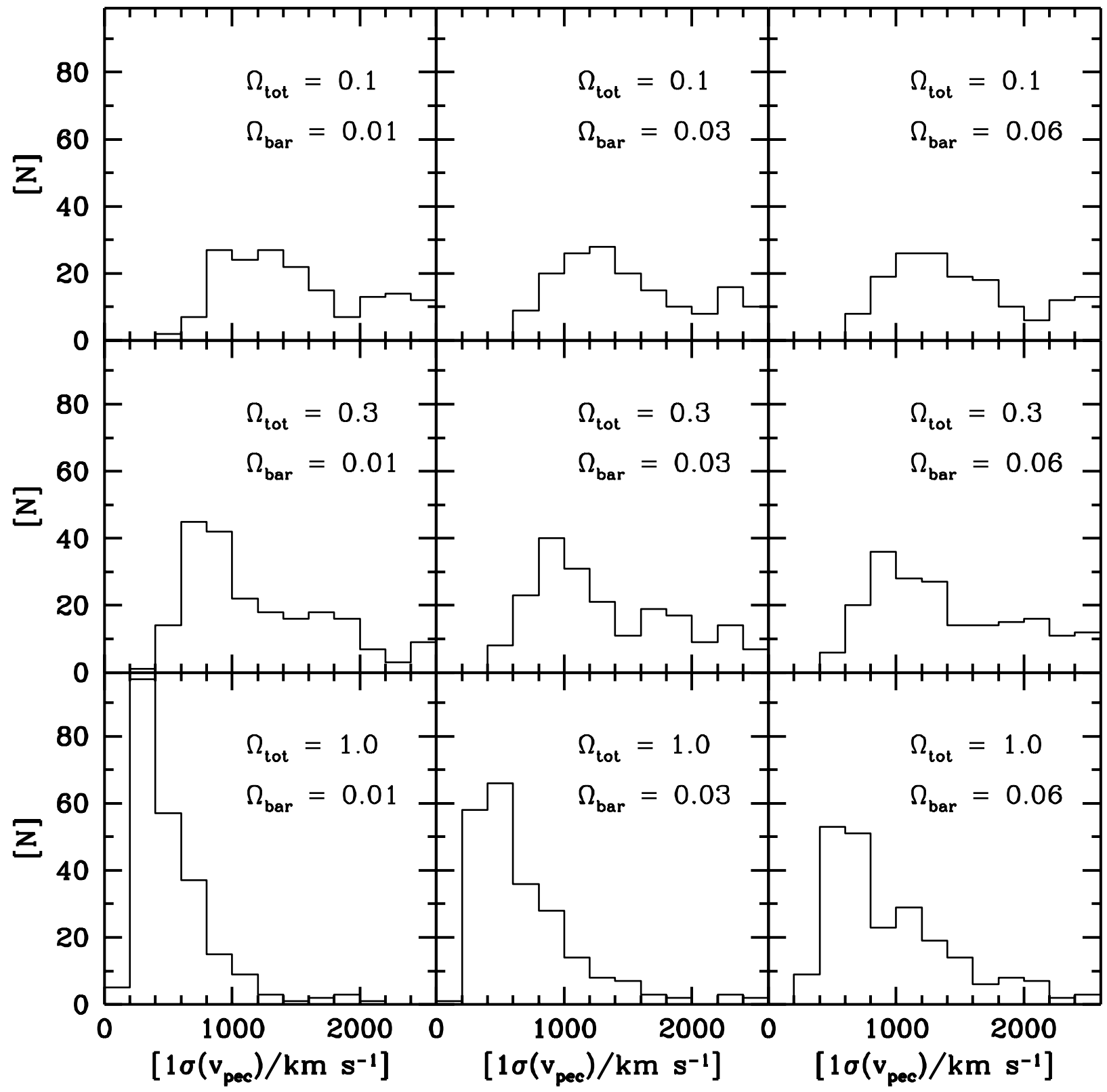

**Figure 7**